# Scalable fabrication of nanostructured p-Si/n-ZnO heterojunctions by femtosecond-laser processing

D.G. Georgiadou[1], M. Ulmeanu[2,3], M. Kompitsas[4], P. Argitis[1], and M. Kandyla[4,*]

[1]National Center for Scientific Research 'Demokritos', Institute of Microelectronics, 15310 Athens, Greece

[2]National Institute for Laser, Plasma and Radiation Physics, Laser Department, Atomistilor 409, P.O. Box MG-36, 077125 Magurele, Romania

[3]University of Bristol, School of Chemistry, Bristol BS8 1TS, UK

[4]National Hellenic Research Foundation, Theoretical and Physical Chemistry Institute, 48 Vasileos Constantinou Avenue, 11635 Athens, Greece

*E-mail: kandyla@eie.gr

## Abstract

We present a versatile, large-scale fabrication method for nanostructured semiconducting junctions. Silicon substrates were processed by femtosecond laser pulses in methanol and a quasi-ordered distribution of columnar nanospikes was formed on the surface of the substrates. A thin (80 nm) layer of ZnO was deposited on the laser-processed silicon surface by pulsed laser deposition, forming a nanostructured p-Si/n-ZnO heterojunction. We characterized the structural, optical, and electrical properties of the heterojunction. Electrical I-V measurements on the nanostructured p-Si/n-ZnO device show non-linear electric characteristics with a diode-like behavior. Electrical I-V measurements on a flat p-Si/n-ZnO reference sample show similar characteristics, however the forward current and rectification ratio are improved by orders of magnitude in the nanostructured device. The fabrication method employed in this work can be extended to other homojunctions or heterojunctions for electronic and optoelectronic devices with large surface area.

# 1. Introduction

ZnO is a promising material for optoelectronic and transparent electronic applications due to its wide band gap (~3.4 eV), efficient ultraviolet (UV) emission ($\lambda$ = 380 nm), and large exciton binding energy (60 meV). Recently, several studies have focused on the development of light-emitting diodes (LEDs) and photodetectors based on ZnO thin films and nanostructures [1,2,3,4]. Nanostructured electronic devices benefit from the advantages of large surface-to-volume ratio and modified light-matter interaction [5,6].

ZnO is naturally an n-type semiconductor due to oxygen vacancies and zinc interstitials [3]. Because of the difficulty in fabricating p-type ZnO, due to low solubility of dopants, most ZnO-based optoelectronic devices rely on heterojunctions between n-type ZnO and p-type semiconducting materials, the most common choice being p-type silicon. Heterojunction p-Si/n-ZnO devices have been employed as UV-visible photodetectors [7,8] and LEDs [9,10,11]. Recently, ZnO nanostructures, such as nanorods, nanodots, and nanowires have been grown in junction with p-Si substrates, for the development of photodetectors [12,13] and LEDs [2,14,15,16]. Various fabrication methods have been employed for the formation of ZnO nanostructures on silicon, including vapor liquid solid growth, aqueous chemical growth, hydrothermal synthesis, and pulsed laser deposition, among others [12,13,15,16]. Additional fabrication steps are usually required in order to deposit metallic contacts on the ZnO nanostructures without creating a short circuit with the silicon substrate, which are often incompatible with large-scale production. The scalability of the fabrication method is an important parameter, affecting the possibility to adapt this new technology to mass production of electronic and photonic circuits.

In this work, we present a simple and scalable fabrication method for nanostructured p-Si/n-ZnO heterojunctions. We employ femtosecond-laser nanostructuring of silicon substrates, on which we deposit a layer of ZnO. We characterize the structural, optical, and electrical properties of the device and we find that a diode has been formed between ZnO and nanostructured silicon, which shows enhanced electrical response, compared with unstructured reference samples. The fabrication method can be easily extended to other homojunctions or heterojunctions with large surface area, for electronic and photonic applications.

## 2. Experimental

Nanostructured silicon substrates were prepared by femtosecond laser irradiation in methanol [17]. A commercial femtosecond Ti:sapphire laser system was employed, with 90 fs pulse duration, 1 kHz repetition rate, $M^2 \sim 1.2$, and peak pulse energy 350 μJ. The laser beam was frequency-doubled by a second harmonic crystal to a wavelength of 400 nm. The beam was focused via a microscope objective lens (Newport; UV Objective Model, 37x, 0.11 NA) onto p-type silicon (111) samples, with a resistivity of 1 – 30 Ω cm, placed at the bottom of an open container, which was filled with methanol. The silicon samples were placed on a computer-controlled XYZ positioning stage from Aerotech (XY stages: 0.1 μm resolution and Z stage: 0.5 μm resolution and ±2 μm repeatability). The laser power on the sample was controlled by the rotation of a half-wave plate that changed the polarization plane of the laser beam before the second harmonic crystal. A circular mask was used in order to remove the tails of the Gaussian profile of the beam. Single-pass grooves were machined on the silicon samples by using different laser power values and different translation speed values. We have used three different translation speeds of 0.1 mm/s,

0.25 mm/s, and 0.5 mm/s, which correspond to a distance between two consecutive laser pulses of 0.1 μm, 0.25 μm, and 0.5 μm, on the silicon surface, respectively. The laser power varied from 1 mW to 20 mW. Once the experimental parameters were optimized for processing one line, surface machining of larger areas, line after line, was possible.

A thin layer of ZnO was deposited on the nanostructured silicon surface by pulsed laser deposition [18,19]. For this purpose, a Q-switched Nd:YAG laser system (355 nm wavelength, 10 ns pulse duration) was employed for the ablation of a Zn target inside a vacuum chamber. The nanostructured silicon sample was placed across the Zn target, at a distance of 45 mm, and was heated to 300°C during thin-film deposition. The chamber was evacuated to a residual pressure of $10^{-5}$ mbar prior to laser irradiation of the Zn target. The deposition took place under an oxygen flow of 20 Pa dynamic pressure, in order for ZnO to form on the nanostructured silicon surface. The Zn target irradiation time was 60 min. A schematic of the final device structure is shown in Fig. 1.

The surface morphology of the samples was characterized by scanning electron microscopy (SEM). Room temperature photoluminescence measurements were performed using a He/Cd laser with an excitation wavelength of 325 nm. Dark I-V characteristics were measured with a Keithley 2400 source-measure unit at room temperature in air. Figure 1 shows the electric contacts employed for I-V measurements. The ZnO layer was grounded and a bias voltage was applied to the silicon substrate (with a native $SiO_2$ layer). Measurements were obtained with tungsten microprobe tips in direct contact with the Si/ZnO surfaces.

## 3. Results and discussion

Figure 2a shows an optical microscope image of the femtosecond-laser nanostructured silicon surface. The area of the structured surface is 1090 μm × 1080 μm, which is conveniently large for applications. The structured area appears black, compared with the unstructured surrounding silicon surface, because it absorbs strongly in the visible [20]. Figures 2b – 2c show SEM images of the femtosecond-laser nanostructured silicon samples, coated by ZnO. We observe that the silicon surface is covered by a quasi-ordered distribution of columnar nanospikes. Femtosecond laser irradiation of silicon in liquid environments is known to generate nanospikes on the surface of the material due to ultrafast melting and interference effects [17,21]. The thickness of the ZnO layer was measured by white light reflectance spectroscopy on a flat silicon substrate, on which ZnO was deposited under identical conditions to those employed for the nanostructured silicon substrate, and was found to be 80 nm. From SEM images of the nanostructured silicon samples before ZnO coating (not shown here), we observe that the ZnO deposition does not change the morphology of the silicon nanospikes. The active surface area of the final p-Si/n-ZnO device is significantly increased, due to the nanostructure of the silicon substrate.

Figure 3 shows the room temperature photoluminescence spectrum of ZnO, deposited on the nanostructured silicon substrate. The strong UV band at 378 nm is characteristic of crystalline ZnO and is attributed to the excitonic emission of the material [22,23]. The weaker broad band around 630 nm originates from defects in the ZnO layer, which produce deep level emissions in the yellow-orange region. The nature of these defects remains an open research topic, however it has been shown that they often consist of extrinsic deep centers, such as Cu, Fe, *etc.*, and intrinsic centers, such as oxygen and zinc interstitials, as well as oxygen and zinc vacancies [3,24,25].

Figure 4a shows the I-V curve of the nanostructured p-Si/n-ZnO device, which is shown in Fig. 1. We observe that the device shows non-linear electric characteristics with a diode-like behavior and a rectification ratio of $I_{forward}/I_{reverse} \cong 1.5 \times 10^3$ at ±20 V. The reverse leakage current is found to be only 0.07 μA at –20 V. Figure 4b shows the I-V curve of a flat p-Si/n-ZnO reference device (inset). For this sample, ZnO was deposited under identical conditions to those employed for the device shown in Fig. 1, on a flat Si substrate, identical to the one employed in Fig. 1. We observe that the reference device also shows a diode-like behavior, with a rectification ratio of $I_{forward}/I_{reverse} \cong 42$ at ±20 V, however the forward current is 2-3 orders of magnitude lower compared to the nanostructured device. The reverse leakage current is found 0.01 μA at –20 V.

Both the flat and the nanostructured devices show a rectifying behavior, indicating a p-n junction has been formed between the p-type silicon substrate and the n-type ZnO layer. We note that an ohmic contact is expected to form between the ZnO layer (work function: 4.4 – 4.7 eV [26]) and the tungsten microprobe tip (work function: 4.5 eV [27]). Therefore, the rectifying behavior of the I-V curves is due to the Si/ZnO p-n junction rather than to the point contacts employed for these measurements. Comparing Figs. 4a and 4b we observe that the nanostructured device shows enhanced electric characteristics, with orders of magnitude improvement in the forward current and the rectification ratio. The improved electric characteristics of the nanostructured device originate from the increased surface area of the p-n junction between the ZnO layer and the silicon nanospikes. The reverse leakage current remains low for both types of devices. The fact that the increase of the forward current is two orders of magnitude higher than the increase of the reverse current in the nanostructured device, can be justified by taking into account the difference in the injecting contact when biasing is reversed. More specifically, the high forward current in the nanostructured device is attributed to the local enhancement of the electric field in the silicon nanospikes (having a positive potential towards the ZnO layer at the interface), something which is not happening when bias is reversed.

The scalability of the fabrication method presented in this work is only limited by the size of the substrate silicon wafer and the travel range of the computer-controlled positioning stages, which translate the silicon substrate with respect to the femtosecond laser beam for nanostructuring. No other process inherently limits the dimensions of the nanostructured silicon area. Once the nanostructured silicon substrate is fabricated, a layer of ZnO can be deposited at the top with a variety of methods, to form a nanostructured heterojunction. Because the ZnO layer completely covers the nanostructured silicon area, a top electrode can be easily deposited on ZnO without special care to avoid creating a short circuit with silicon. The addition of metallic contacts is expected to increase further the electric current extracted from the devices [28]. The silicon substrate can also be combined with a number of other layers, in order to form a variety of nanostructured homojunctions or heterojunctions.

**4. Conclusions**

We fabricated a nanostructured p-Si/n-ZnO heterojunction and characterized its structural, optical, and electrical properties. Silicon substrates were processed by a femtosecond laser in methanol. A quasi-ordered distribution of columnar nanospikes was formed on the surface of the substrates. A thin layer of ZnO was deposited on the nanostructured silicon surface by pulsed laser deposition. Photoluminescence measurements on ZnO show a strong UV band at 378 nm, which is attributed to the excitonic emission of the material, and a weaker broad band around 630 nm, which originates from defects in the ZnO structure. Electrical I-V measurements on the nanostructured p-Si/n-ZnO device show non-linear, diode-like characteristics. Electrical I-V measurements on a flat p-Si/n-ZnO reference device show similar characteristics, however the forward current and rectification ratio are improved by orders of magnitude in the nanostructured device. The fabrication method employed in this work is scalable and versatile, and can be extended to other homojunctions or heterojunctions for electronic and optoelectronic devices with large surface area.

**Acknowledgments**

The project was funded by the John S. Latsis Public Benefit Foundation. The sole responsibility for the content lies with its authors. M. Ulmeanu is supported by a Marie Curie International Intra-European Fellowship (FP7-PEOPLE-2013-IEF-625403). The authors would like to thank Dr. N. Boukos from the Institute of Materials Science, NCSR 'Demokritos', for photoluminescence measurements.

**Figure captions**

Figure 1: Schematic representation of the nanostructured p-Si/n-ZnO device.

Figure 2: (a) Optical microscope image of the femtosecond-laser nanostructured silicon surface and (b), (c) SEM images (viewed at 25°) of the same sample coated with ZnO.

Figure 3: Photoluminescence spectrum of ZnO deposited on the nanostructured silicon substrate, taken at room temperature in air with a He/Cd laser, operating at 325 nm.

Figure 4: (a) I-V curve of the nanostructured device shown in Fig. 1. (b) I-V curve of a flat reference device (inset).

**Figures**

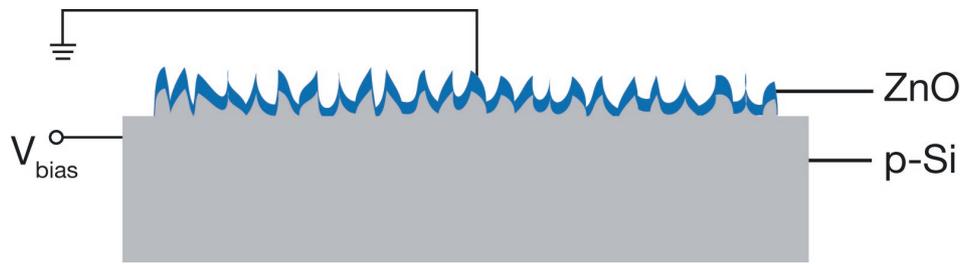

Figure 1

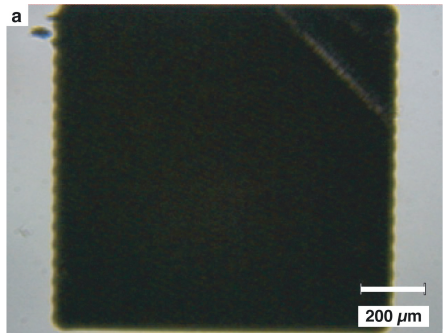

Figure 2a

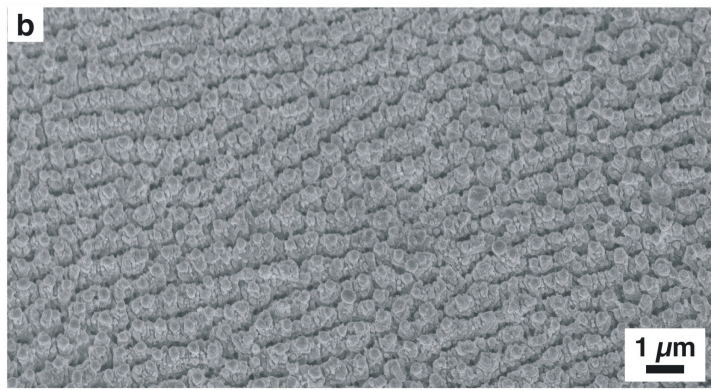

Figure 2b

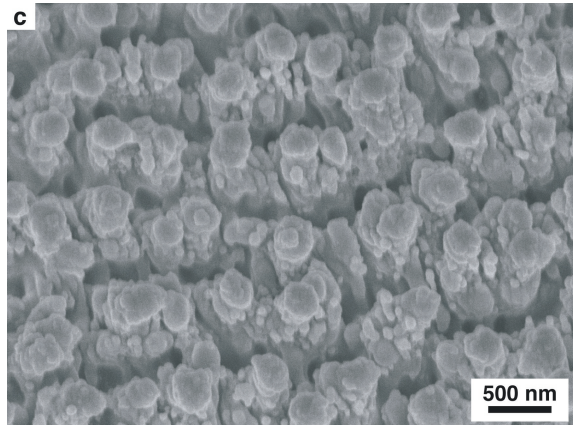

Figure 2c

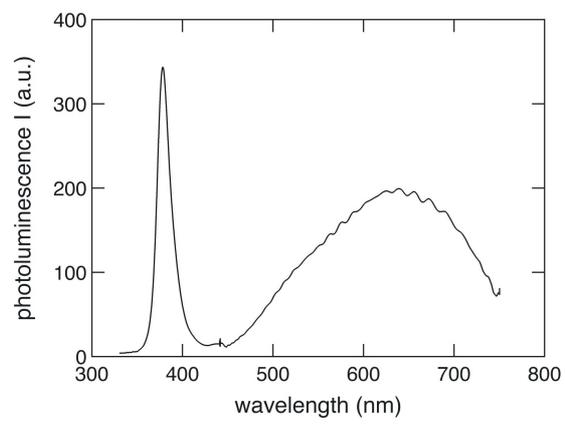

Figure 3

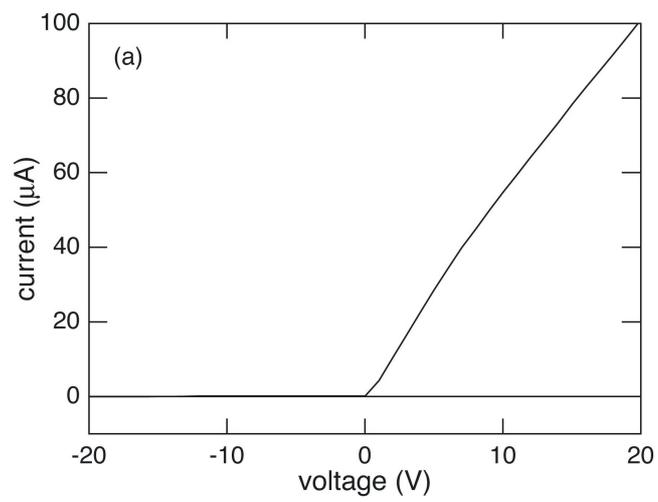

Figure 4a

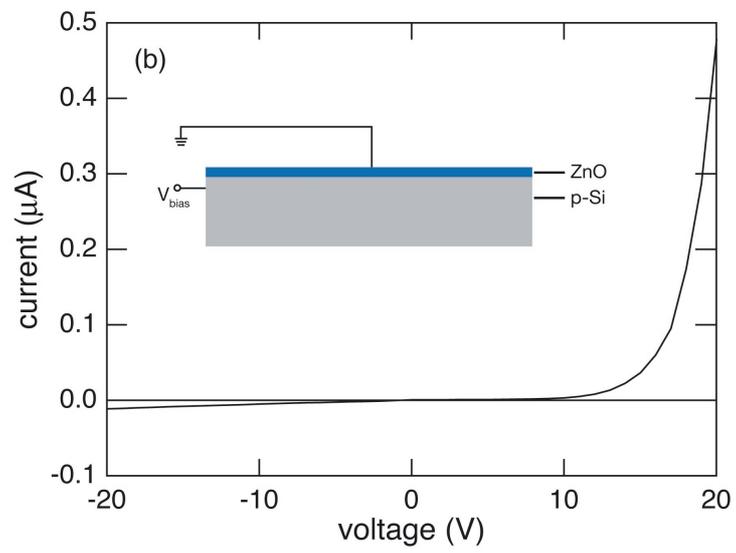

Figure 4b